# Phase retrieval without prior knowledge via single-shot Fraunhofer diffraction pattern of complex object

An-Dong Xiong[1], Xiao-Peng Jin[1], Wen-Kai Yu[1] and Qing Zhao[1*]

**Fraunhofer diffraction is a well-known phenomenon achieved with most wavelength even without lens. A single-shot intensity measurement of diffraction is generally considered inadequate to reconstruct the original light field, because the lost phase part is indispensable for reverse transformation. Phase retrieval is usually conducted in two means: priori knowledge or multiple different measurements. However, priori knowledge works for certain type of object while multiple measurements are difficult for short wavelength. Here, by introducing non-orthogonal measurement via high density sampling scheme, we demonstrate that one single-shot Fraunhofer diffraction pattern of complex object is sufficient for phase retrieval. Both simulation and experimental results have demonstrated the feasibility of our scheme. Reconstruction of complex object reveals depth information or refraction index; and single-shot measurement can be achieved under most scenario. Their combination will broaden the application field of coherent diffraction imaging.**

Fraunhofer diffraction, also known as far-field diffraction, does not necessarily need any extra optical devices except a beam source. The diffraction field is the Fourier transform of the original light field. However, for visible light or X-ray diffraction, the phase part is hardly directly measurable[1]. Therefore, phase retrieval from intensity measurement becomes necessary for reconstructing the original field. In a broad sense, there are two ways to compensate for the lost phase part. One way is adding prior knowledge about the object; another way is taking multiple different measurements. In 1978, Fienup developed an algorithm based on alternating projection using nonnegativity as a constraint of the recovered image[2,3]. Reconstruction for noncrystalline object from the intensity measurement of Fraunhofer diffraction was achieved by Miao et al. in 1999[4,5]. With the development of compressive sensing theory, sparsity of the image in certain transformation is also incorporated to solve phase retrieval problem[6-9]. Recently, method called PhaseMax is raised pursuing the solution maximizing the inner product with an approximation vector[10,11]. These approaches above use prior knowledge to make up for the lost phase part. On the other hand, if several different measuring matrixes can be cast on the diffraction setups, matrix completion methods like Phase Lift and Wirtinger Flow can perform well even for complex problem[12-14]. These matrix complement methods require 4N-4 (N stands for the dimension) generic measurements such as Gaussian random measurements for complex field[15]. Ptychography is another reliable method to achieve image with good resolution if the sample can endure scanning[16-19]. To achieve higher spatial resolution around the size of atom, beams with shorter wavelength like hard X-ray or electrons are widely used these days. But these high energy particles will heat the sample or even break the chemical bonds[20]. Multiple measurements (like ptychography or matrix complement methods) appear difficult to be implemented for fragile sample under hard beams. Both priori knowledge and multiple measurements have their advantages in different circumstance if the object can meet certain condition. However, sometimes conflicts occur between our needs and demand of measuring methods. For example, getting 3D information of biological sample 'alive' with high resolution means complex object and only one beam shot with few optical devices. The most demanding phase retrieval shows little noise endurance and usually ends up in simulation. Application of phase retrieval is usually conducted by assuming the object is real, even when the diffraction pattern is non-centrosymmetric (meaning complex object). Here, we demonstrate the feasibility of reconstructing complex object without prior information from the intensity measurement of single-shot Fraunhofer diffraction in lab condition.

[1] Center for Quantum Technology Research, School of Physics, Beijing Institute of Technology, Beijing 100081, China
* Email: qzhaoyuping@bit.edu.cn

The light field of two-dimentional Fraunhofer diffraction can be computed by

$$F(\theta,\varphi) = \iint dx_1 dx_2 O(x_1,x_2) e^{i\frac{2\pi}{\lambda}(x_1\cos\theta\cos\varphi + x_2\cos\theta\sin\varphi)}$$

Here, $\vec{k} = (\cos\theta\cos\varphi, \cos\theta\sin\varphi, \sin\theta)$ represents the unit vector of emergent light; $\lambda$ is wavelength; $O(x_1,x_2)$ denotes the light field of the illuminated object. Fast Fourier transform (FFT), whose bases are orthogonal and can be quickly computed, is the most widely used discrete Fourier transform. Due to independence of these measurement vectors, if not provided with any prior information, any phase multiplied by the intensity of diffraction pattern for complex original field is reasonable and can be inverse transformed. To constraint the phase part, we need non-orthogonal measurement vectors, which means higher sampling density in $k$ domain (diffraction plane). In order to distinguish from FFT, we call this scheme high-density Fourier transform (HFT).

If the diffraction field is measured completely including the phase, the measurement can be taken as a linear process, multiplying vectorized signal of original light field $|o_x\rangle$ by measurement matrix $M$. Measurement $F_k$ at different position $k$ of diffraction plane is the inner product of corresponding row $\langle m_k|$ and $|o_x\rangle$.

$$F_k = \langle m_k | o_x \rangle = \sum_j \langle m_k | m_j \rangle \langle m_j | o_x \rangle = \sum_j P_{kj} F_j$$

, where $|m_j\rangle$ is a set of complete orthogonal basis.

The inner product $P_{\alpha\beta}$ of the measurement of $k_\alpha$ and $k_\beta$ is given by

$$P_{\alpha\beta} = \frac{1}{l_1 l_2} \int_{w_2}^{w_2+l_2} dx_2 \int_{w_1}^{w_1+l_1} dx_1 e^{\frac{2\pi i}{\lambda}[x_1(k_{\alpha1}-k_{\beta1})+x_2(k_{\alpha2}-k_{\beta2})]}$$

$$= \frac{e^{\frac{2\pi i}{\lambda}(k_{\alpha1}-k_{\beta1})(w_1+l_1)} - e^{\frac{2\pi i}{\lambda}(k_{\alpha1}-k_{\beta1})w_1}}{\frac{2\pi i l_1}{\lambda}(k_{\alpha1}-k_{\beta1})} \cdot \frac{e^{\frac{2\pi i}{\lambda}(k_{\alpha2}-k_{\beta2})(w_2+l_2)} - e^{\frac{2\pi i}{\lambda}(k_{\alpha2}-k_{\beta2})w_2}}{\frac{2\pi i l_2}{\lambda}(k_{\alpha2}-k_{\beta2})}$$

$$|P| = \left| \frac{\sin[\frac{\pi l_1}{\lambda}(k_{\alpha1}-k_{\beta1})]}{\frac{\pi l_1}{\lambda}(k_{\alpha1}-k_{\beta1})} \cdot \frac{\sin[\frac{\pi l_2}{\lambda}(k_{\alpha2}-k_{\beta2})]}{\frac{\pi l_2}{\lambda}(k_{\alpha2}-k_{\beta2})} \right|,$$

where $w_{1,2}$ and $w_{1,2}+l_{1,2}$ represent boundary of the original field, $k_{\alpha1,2}$ and $k_{\beta1,2}$ are the horizon and vertical component of $k_\alpha$ and $k_\beta$. Specially, when

$$w_1 = -\frac{1}{2}l_1 \text{ and } w_2 = -\frac{1}{2}l_2,$$

$$P = \frac{\sin[\frac{\pi l_1}{\lambda}(k_{\alpha1}-k_{\beta1})]}{\frac{\pi l_1}{\lambda}(k_{\alpha1}-k_{\beta1})} \cdot \frac{\sin[\frac{\pi l_2}{\lambda}(k_{\alpha2}-k_{\beta2})]}{\frac{\pi l_2}{\lambda}(k_{\alpha2}-k_{\beta2})} \in \mathbb{R}$$

. For convenience, we set $l_1 = l_2 = l$ below.

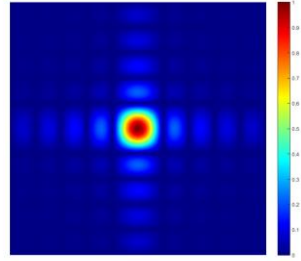

**Fig.1 | Inner product of measuring vectors.** Modulus of inner product for $k_\alpha = (0,0)$, $k_\beta \in [(-\frac{1}{20},-\frac{1}{20}),(\frac{1}{20},\frac{1}{20})]$, $\frac{l}{\lambda} = 100$. $k_\alpha$ is in the center, inner product with itself is 1, while inner product with

$k_\beta = (n_1 * \frac{l}{\lambda}, n_2 * \frac{l}{\lambda})$ ($n_1 \in \mathbb{Z}^*$ or $n_2 \in \mathbb{Z}^*$) is 0.

HFT contains more information than FFT in intensity measurement. Take the following matrix

$$\begin{pmatrix} -0.2515-0.0404i & 0.1371-0.3740i & -0.1159-0.1448i \\ 0.6229-1.1616i & 0.0090-0.3990i & 0.0460-0.2487i \\ 0.1437-0.3831i & 0.0295-0.3449i & 0.9242-1.6586i \end{pmatrix}$$

for example, the magnitude of shifted (shifting the zero-frequency component to the center) FFT $\begin{pmatrix} 3 & 2 & 1 \\ 1 & 5 & 1 \\ 1 & 2 & 3 \end{pmatrix}$ is centrosymmetric, while the 3×3 times HFT

$$\begin{pmatrix} 2.0068 & 2.8691 & 2.5062 & 1.3127 & 1.4950 & 2.6203 & 2.7675 & 1.7153 & 0.5123 \\ 2.8418 & 3 & 1.8934 & 5.789 & 2 & 2.8311 & 2.3641 & 1 & 1.5491 \\ 2.8992 & 2.4411 & 0.8247 & 1.2582 & 2.8169 & 3.1796 & 2.2561 & 1.0879 & 2.0708 \\ 2.6019 & 1.5598 & 0.5990 & 2.7811 & 4.0811 & 3.8968 & 2.4017 & 1.0879 & 2.1744 \\ 2.5966 & 1 & 1.6632 & 3.9892 & 5 & 4.2248 & 2.0843 & 1 & 2.5267 \\ 2.7206 & 0.8689 & 2.3112 & 4.4093 & 4.9224 & 3.5764 & 1.0408 & 1.7153 & 3.1297 \\ 2.3040 & 0.5789 & 2.5272 & 3.9595 & 3.7562 & 1.9958 & 0.5767 & 2.6218 & 3.3089 \\ 1.1348 & 1 & 2.5836 & 2.9810 & 2 & 0.4123 & 1.9560 & 3 & 2.6749 \\ 0.5466 & 2.0701 & 2.6465 & 2.0272 & 0.9139 & 1.7703 & 2.7543 & 2.6218 & 1.3625 \end{pmatrix}$$

is not; because the magnitude part of orthogonal integer measurement is centrosymmetric while the phase part is not. The fractional measurement, linear superposition of integer measurement, can reveal phase information via intensity. Phase of Fourier transform is usually believed to carry more information; the image of

$O_{mix} = iFFT(|FFT(O_1)| * \frac{FFT(O_2)}{|FFT(O_2)|})$ is visually more like $O_2$. But in the case of HFT, we can see the overlap of two set of conjugate phase solution of $O_1$ (complex conjugate in $k$ domain means inversion of complex conjugate in $x$ domain) without any iteration algorithm from the image of $O_{mix} = iHFT(|HFT(O_1)| * \frac{HFT(O_2)}{|HFT(O_2)|})$, where $O_2$ is an all-ones matrix. As can be seen from Fig.2b, shape of $O_1$ emerges at high sampling ratio $R = r^2$ for certain noise level, where $\frac{1}{r} \cdot \frac{\lambda}{l}$ is the interval between adjacent sampled measurement. Intensity of Fraunhofer diffraction is usually in dramatically uneven distribution, different types of noise have different effect on performance of phase retrieval. Noise in simulation hereinafter is in the form: $a' = a * rnoi * randn + anoi * randn$, where $a$ is the magnitude and $randn$ is a Gaussian random matrix generated by Matlab whose variance is 1. To some extent, in order to retrieve the complex field from single-frame noisy Fraunhofer diffraction, the sampling ratio needed should be much higher than the regular 4× or 8×.

As shown in Fig.1, the inner products exhibit strong locality. They are zero at "integer" positions, which correspond to the basis of FFT. The superposition coefficient will change from $P_{kj}$ to $P_{k'j} = P_{kj} + \Delta P_{kk'j}$ when $k \to k' = k + \frac{1}{r} \cdot \frac{\lambda}{l}$. $\Delta P_{kk'j}$ is much larger for neighbor 'integer' measurement than non-neighbor ones. With higher sampling density, non-neighbor coefficient difference can be ignored, making the quadratic problem more localized and provide some robustness. Still, for the vast magnitude range of Fourier transform, even hundred times coefficient may not be good enough for high noise measurement. A quadratic problem like $|p_1 f_1 + p_2 f_2| = y_3$, $|f_1| = y_1$, $|f_2| = y_2$ will produce two conjugate solutions of phase difference. Every adjacent pair will give two solutions and the combination will be $2^{N-1}$ if coefficient between measurement with $k$ distance more than $\frac{\lambda}{l}$ is ignored. Therefore, phase retrieval without other constraint is difficult for 1D problem. For 2D case, four points will form a loop and the phase $\sigma$ may change from some value $\sigma_0$ to the value $\sigma_0 + 2\pi n$ ($n$ is an integer) as going around the loop to the same starting point. Loop constraint for all loops in multi-dimensional case is equivalent to loop constraint for all 4-point loops (see supplementary information). Without noise, only two conjugate solutions will be generated under these loop constraints.

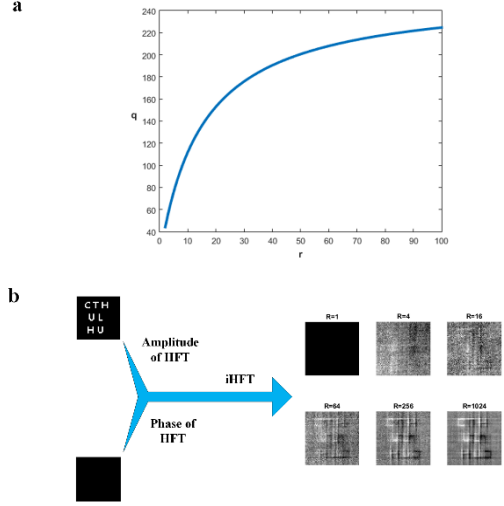

**Fig.2 | Demonstration of different sampling density. a:** coefficient ratio between adjacent and secondary adjacent measurement by sampling density. Where $q = \frac{\Delta P_{kk'j_1}}{\Delta P_{kk'j_2}}$, $k = \frac{1}{20} \cdot \frac{\lambda}{l}$, $k' = k + \frac{1}{r} \cdot \frac{\lambda}{l}$, $j_1 = 0$, $j_2 = -1$ **b:** Inverse HFT of phase multiplication for different sampling ratio. Images demonstrated are the imaginary part. Values of "CTHULHU" are all $i$, while those of the background are all 1. Here we set $rnoi = 0.5$, $anoi = 100$. Mean value of imaginary part on the right is 0.

For discrete case, though the phase part is different from continuous case (see supplementary information), HFT can be computed by

$$F(k_1, k_2) = \sum_{n_{1,2}=1}^{N} O(n_1, n_2) e^{\frac{i2\pi n_1 k_1}{N_1} + \frac{i2\pi n_2 k_2}{N_2}},$$

$$(k_{1,2} = \frac{1}{r}, \frac{2}{r}, ..., \frac{N_{1,2} \times r}{r})$$

which can be converted into

$$F(k_1', k_2') = \sum_{n_{1,2}=1}^{N \times r} O'(n_1, n_2) e^{\frac{i2\pi n k_1'}{N_1 \times r} + \frac{i2\pi n k_2'}{N_2 \times r}}$$

$k_{1,2}' = 1, 2, ..., N_{1,2} \times r$

$$O'(n_1, n_2) = \begin{cases} O(n_1, n_2) & n_{1,2} \in G_1 = [1, N_{1,2}] \\ 0 & n_{1,2} \in G_2 = [N_{1,2} + 1, N_{1,2} \times r] \end{cases}$$

The intensity computed by this method is consistent with experiment result (see Fig.3).

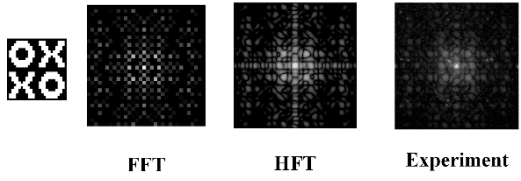

| FFT | HFT | Experiment |

**Fig.3 | Magnitude of simulation and experiment for FFT and HFT of XO.** Black part of XO is 1, white part is i. HFT here is computed in converted method.

Via this conversion, measuring procedure can be fast computed and the quadratic problem is turned into pursuing zero in the zero zone $G_2$, compatible with many existing algorithms. Here we use the classic hybrid input-output (HIO) algorithm[3], which is widely used in coherent diffraction imaging (illustrated in Fig.4):

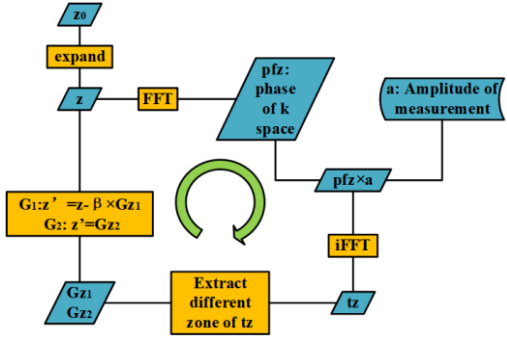

**Fig.4 | Diagram of HIO algorithm.** Procedure of expanding depends on $G_1$ and size of measurement. β can be a partially random matrix.

Input: $a$, $T$, $tol$

  $a$- magnitude of Fraunhofer diffraction measurement

  $T$- max number of iteration

  $tol$- toleration for sum of $G_2$

Output: $z$-reconstructed original field

Initialization: choose initial $z_0$ (e.g. a random or all-ones matrix),

$$z = \begin{cases} z_0(n_1,n_2) & n_{1,2} \in G_1 \\ 0 & n_{1,2} \in G_2 \end{cases}$$

For $i = 1, 2, \ldots$

(1) $pfz = \dfrac{FFT(z)}{|FFT(z)|}$ ;

(2) $tz = iFFT(pfz * a)$, where $*$ represents Hadamard multiplication;

(3) $gz(n_1,n_2) = \begin{cases} 0 & n_{1,2} \in G_1 \\ tz(n_1,n_2) & n_{1,2} \in G_2 \end{cases}$

(4) $z(n_1,n_2) = \begin{cases} tz(n_1,n_2) & n_{1,2} \in G_1 \\ z(n_1,n_2) - \beta \bullet gz(n_1,n_2) & n_{1,2} \in G_2 \end{cases}$

Until: $i = T$ or $S = \sum_{n_{1,2} \in G_2} |z(n_1,n_2)|^2 \leq tol$

For noise-free or low-noise measurement, reconstruction can be attained even when the original field has random magnitude and phase distribution. Though reconstruction degrades when noise increases, the degradation is mainly caused by overlap of conjugate solutions[26] as indicated in Fig.5d. Noise exceeding superposition coefficient will make $S$ for the overlap of two eigensolutions close to $S$ of the two eigensolutions. It is hard to distinguish the eigensolution from the overlap in the absence of other constraints. To acquire a reconstruction with less overlap, noise level should be controlled near or below superposition coefficient. For noise-free measurement, oversized $G_1$ leads to overlap of solution caused by translation solutions while narrow $G_1$ provides insufficient freedom. Inappropriate zone division will cause rise in $S$, as shown in Fig.5e. For noisy measurement, influence on $S$ of overlap of translation solution can be compared with influence of noise; while degree of freedom has a greater influence on finding solution for noisy measurement. When $G_1$ is too small, S still decreases dramatically with the increasing size of $G_1$; when $G_1$ is oversized, variation of $S$ becomes relatively stable. Therefore, we can determine appropriate size of $G_1$ according to $S$, even if the right size is not provided beforehand.

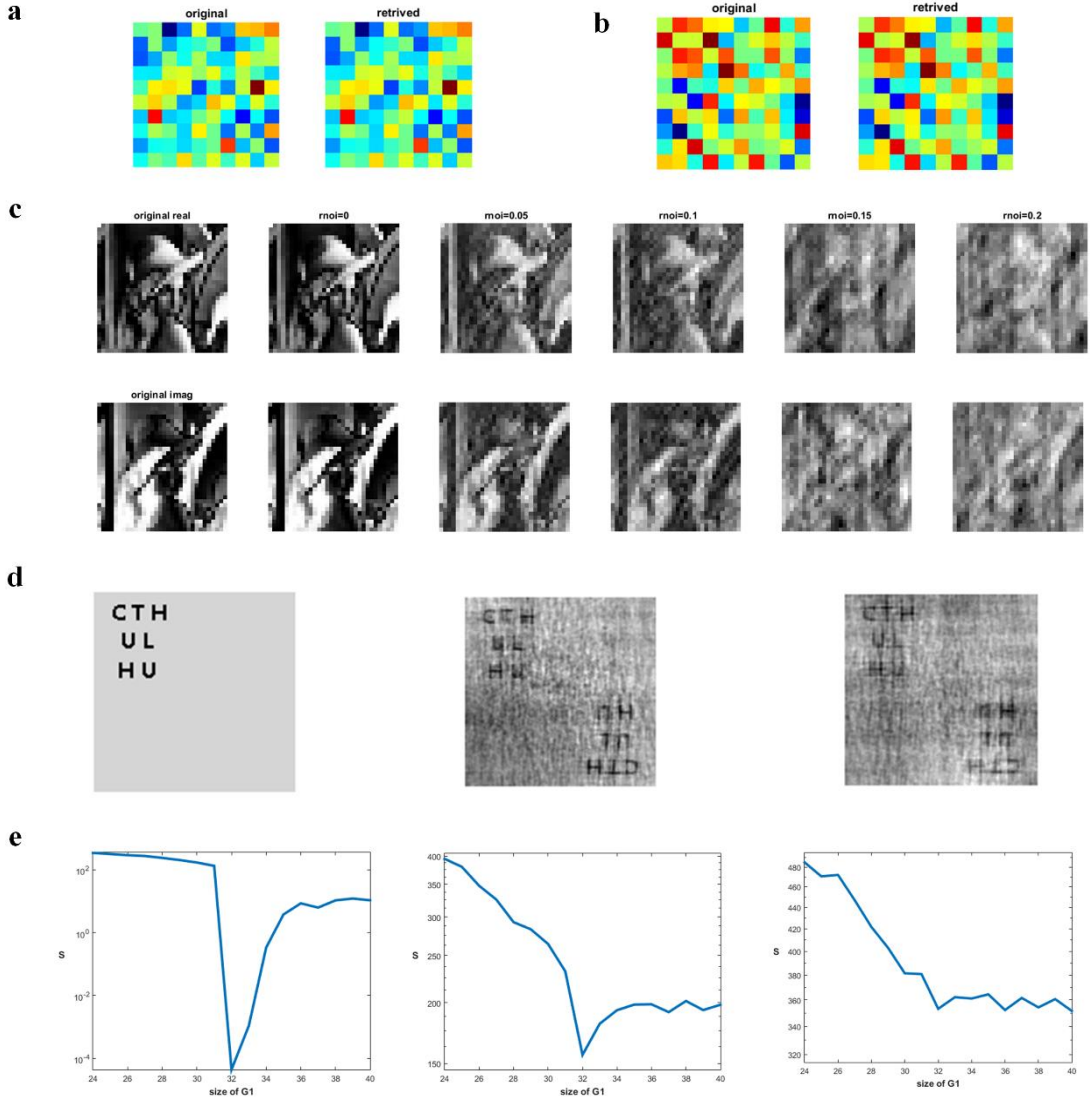

**Fig.5 | Simulation result of phase retrieval based on HFT. Inverted when conjugate solution achieved. a, b**: reconstruction for random complex case, $R=400$, $T=1000$, imaginary part shown. $rnoi=0$ in **a**, while $rnoi=0.1$ in **b**. **c**: reconstruction for $O=e^{2\pi i \frac{Lena}{\max(Lena)}}$, $R=1024$, $T=1000$. **d**: reconstruction for the 'CTHULHU' on the left, $rnoi=0.3, 0.5$, $R=100$, $T=1000$. **e:** logarithmic display of averaged $S$ from 20 runs in reconstruction for $O=e^{2\pi i \frac{Lena}{\max(Lena)}}$, $rnoi=0, 0.1, 0.2$, $R=100$, $T=1000$, the original size is 32×32.

The scheme in experiment setup is depicted in Fig.6. Illuminating beam is generated by 632.8nm He-Ne laser (Thorlabs HNL150RB) and polarized by linear polarizer (Thorlabs LPVISB100-MP2). By applying voltage on liquid crystal on silicon (LCoS) spatial light modulator (SLM) (Meadowlark Optics 1920×1152, pixel pitch 9.2 μm× 9.2 μm), we form complex object with certain phase distribution. Several absorptive neutral density filters (Thorlabs NEK01S) are placed obliquely between optics to reduce secondary reflection, while those vertical to the beam are for adjusting laser intensity. Mapping between k domain and camera pixel is attained by inputting the phase (in 8-bit) of Fourier transform of designed image and locating the known feature points. Pattern in $k$ domain can be shrunk by magnifying image on SLM (see supplementary information). Measurements for reconstruction are all obtained from experiment without any patching with the help of prior information, even for the central part where overexposure might take place.

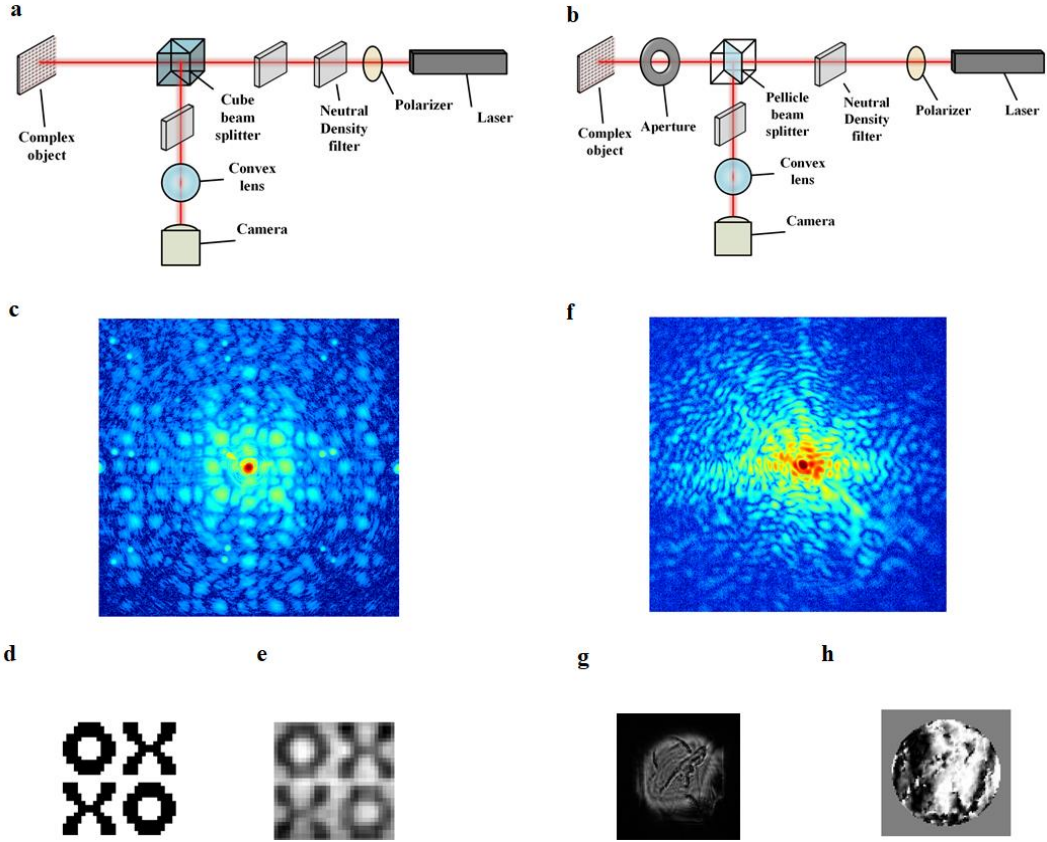

**Fig.6 | Experiment setup and result. a, b**: diagram and of experiment setup. **a** is for "OXXO" and **b** is for "Lena". Neutral density filters are not always in the number shown in **a**, because we have to choose different combination to attain proper intensity of laser. **c**: logarithmic display of measurement collected by CCD camera for diffraction of "OXXO", counting between background noise and overexposure is adopted. **d**: image we cast on SLM. **e**: reconstruction from **c**, real part of phase shown. **f**: logarithmic display of picture captured by 16-bit CMOS camera for diffraction of "Lena". **g**: picture captured by monitoring camera. Due to overlap of point spread function, dark fringe occurs where phase changes rapidly[30]. It is not illuminated in circle because beam is in slight skew and passes the aperture twice to reach the monitoring camera. Diffraction pattern here is mainly caused by aperture. **h**: reconstruction from **f**, real part of phase shown.

For 32×32 "OXXO" image, aperture and beam expander are not employed. Neutral density filters with optical density (OD, transmissivity is $10^{-OD}$) from 0.1 to 4 are placed in various combination, enabling each part of diffraction properly detected by CCD camera (Point Grey GS3-U3-28S4M-C, 1928×1448, 8-bit). The 640×640 measurement is generated from 1353×1353 efficient zone of CCD by averaging vicinal pixels. Star-like noise pixels in measurement with high OD number are eliminated once found too bright than adjacent pixels; because diffraction pattern should be continuous if sampled with high density as ensured by linear superposition coefficient. The algorithm runs with squared $G_I$ and all-ones matrix initial guess. With high sampling ratio R=400, we acquire the reconstruction shown in Fig.6e.

For the 64×64 Lena image, a round aperture and a 10× beam expander (Edmund 55-578) are employed. We use pellicle beam splitters (Thorlabs CM1-BP145B2) instead of cube ones to avoid reflection between inner surfaces. Diffraction pattern is recorded by CMOS camera (Quantalux CS2100M-USB, 1920×1080, 16-bit) in one shot without combination of neutral density filters. The 686×686 measurement is cut directly from the photograph. The algorithm runs with circular $G_I$ and all-ones matrix initial guess. With the sampling ratio R=116, we achieve reconstruction shown in Fig.6h. The reconstruction for asymmetric Lena in this case appears to be asymmetric and preserve some features. Our experiment demonstrates that it is possible to suppress noise level around superposition coefficient and acquire reconstruction with distinguishable overlap.

In conclusion, we demonstrate theoretically that one-shot measurement with high density in *k* domain contains enough information to reconstruct complex original field without prior knowledge. Algorithmically, reconstruction performs well for low-noise measurement, even for random complex object. While in high-noise environments, the degradation of the results is mainly manifested as the superposition of conjugate solutions. Our experiments demonstrate the feasibility of high density sampling and phase retrieval from noisy measurement. We successfully acquire the reconstruction tending to one side of conjugate solutions by suppressing noise adequately.

Fraunhofer diffraction imaging for complex object could help biological real time non-staining microscopy. Future studies may focus on coherent light of hard X-ray wavelength for retrieving phase of field revealing depth and optical features with simplest light path. More techniques in computational imaging can be incorporated into our compatible scheme to avoid overlap and increase noise robustness. Better designed structural phase might exist so that the original field can be obtained directly by multiplying the intensities of high sampling rate without using any iterative algorithms.

## References

1. Shechtman Y, Eldar Y C, Cohen O, et al. Phase retrieval with application to optical imaging: a contemporary overview[J]. IEEE signal processing magazine, 2015, 32(3): 87-109.
2. Fienup J R. Reconstruction of an object from the modulus of its Fourier transform[J]. Optics letters, 1978, 3(1): 27-29.
3. Fienup J R. Phase retrieval algorithms: a comparison[J]. Applied optics, 1982, 21(15): 2758-2769.
4. Miao J, Sayre D, Chapman H N. Phase retrieval from the magnitude of the Fourier transforms of nonperiodic objects[J]. JOSA A, 1998, 15(6): 1662-1669.
5. Miao J, Charalambous P, Kirz J, et al. Extending the methodology of X-ray crystallography to allow imaging of micrometre-sized non-crystalline specimens[J]. Nature, 1999, 400(6742): 342.
6. D.L. Donoho. Compressed sensing[J]. IEEE Transactions on Information Theory, 2006, 52(4):1289-1306.
7. Ohlsson H, Yang A, Dong R, et al. CPRL--An Extension of Compressive Sensing to the Phase Retrieval Problem[C]//Advances in Neural Information Processing Systems. 2012: 1367-1375.
8. Shechtman Y, Beck A, Eldar Y C. GESPAR: Efficient phase retrieval of sparse signals[J]. IEEE transactions on signal processing, 2014, 62(4): 928-938.
9. Sidorenko P, Kfir O, Shechtman Y, et al. Sparsity-based super-resolved coherent diffraction imaging of one-dimensional objects[J]. Nature communications, 2015, 6: 8209.
10. Goldstein T, Studer C. Convex phase retrieval without lifting via PhaseMax[C]//Proceedings of the 34th International Conference on Machine Learning-Volume 70. JMLR. org, 2017: 1273-1281.
11. Goldstein T, Studer C. Phasemax: Convex phase retrieval via basis pursuit[J]. IEEE Transactions on Information Theory, 2018, 64(4): 2675-2689.
12. Candes E J, Strohmer T, Voroninski V. Phaselift: Exact and stable signal recovery from magnitude measurements via convex programming[J]. Communications on Pure and Applied Mathematics, 2013, 66(8): 1241-1274.
13. Candes E J, Eldar Y C, Strohmer T, et al. Phase retrieval via matrix completion[J]. SIAM review, 2015, 57(2): 225-251.
14. Candes E J, Li X, Soltanolkotabi M. Phase retrieval via Wirtinger flow: Theory and algorithms[J]. IEEE Transactions on Information Theory, 2015, 61(4): 1985-2007.
15. Bandeira A S, Cahill J, Mixon D G, et al. Saving phase: Injectivity and stability for phase retrieval[J]. Applied and Computational Harmonic Analysis, 2014, 37(1):106-125.
16. Rodenburg J M. Ptychography and related diffractive imaging methods[J]. Advances in imaging and electron physics, 2008, 150: 87-184.
17. Tian L, Li X, Ramchandran K, et al. Multiplexed coded illumination for Fourier Ptychography with an LED array microscope[J]. Biomedical optics express, 2014, 5(7): 2376-2389.
18. Holler M, Díaz A, Guizar-Sicairos M, et al. X-ray ptychographic computed tomography at 16 nm isotropic 3D resolution[J]. Scientific reports, 2014, 4: 3857
19. Pfeiffer F. X-ray ptychography[J]. Nature Photonics, 2018, 12(1): 9-17.
20. Sayre D, Chapman H N. X-ray microscopy[J]. Acta Crystallographica Section A: Foundations of Crystallography, 1995, 51(3): 237-252.
21. Gauthier D, Guizar-Sicairos M, Ge X, et al. Single-shot femtosecond X-ray holography using extended references[J]. Physical review letters, 2010, 105(9): 093901.
22. Loh N D, Hampton C Y, Martin A V, et al. Fractal morphology, imaging and mass spectrometry of single aerosol particles in flight[J]. Nature, 2012, 486(7404): 513.
23. Starodub D, Aquila A, Bajt S, et al. Single-particle structure determination by correlations of snapshot X-ray diffraction patterns[J]. Nature communications, 2012, 3: 1276.
24. Gallagher-Jones M, Bessho Y, Kim S, et al. Macromolecular structures probed by combining single-shot free-electron laser diffraction with synchrotron coherent X-ray imaging[J]. Nature communications, 2014, 5: 3798.
25. Zhang F, Chen B, Morrison G R, et al. Phase retrieval by coherent modulation imaging[J]. Nature communications, 2016, 7: 13367.
26. Guizar-Sicairos M, Fienup J R. Understanding the twin-image problem in phase retrieval[J]. Journal of the Optical Society of America A, 2012, 29(11):2367.
27. Barty A, Boutet S, Bogan M J, et al. Ultrafast single-shot diffraction imaging of nanoscale dynamics[J]. Nature Photonics, 2008, 2(7): 415.

28. Gardner D F, Tanksalvala M, Shanblatt E R, et al. Subwavelength coherent imaging of periodic samples using a 13.5 nm tabletop high-harmonic light source[J]. Nature Photonics, 2017, 11(4): 259.

29. Song C, Jiang H, Mancuso A, et al. Quantitative imaging of single, unstained viruses with coherent x rays[J]. Physical review letters, 2008, 101(15): 158101.

30. Yu W K, Xiong A D, Yao X R, et al. Efficient phase retrieval based on dark fringe recognition with an ability of bypassing invalid fringes[J]. arXiv preprint arXiv:1612.04733, 2016.

31. Robinson I K, Vartanyants I A, Williams G J, et al. Reconstruction of the shapes of gold nanocrystals using coherent x-ray diffraction[J]. Physical review letters, 2001, 87(19): 195505.

32. Duarte J, Cassin R, Huijts J, et al. Computed stereo lensless X-ray imaging[J]. Nature Photonics, 2019: 1.

33. Rodriguez J A, Xu R, Chen C C, et al. Oversampling smoothness: an effective algorithm for phase retrieval of noisy diffraction intensities[J]. Journal of applied crystallography, 2013, 46(2): 312-318.

34. Marchesini S, He H, Chapman H N, et al. X-ray image reconstruction from a diffraction pattern alone[J]. Physical Review B, 2003, 68(14): 140101.

35. Chapman H N, Fromme P, Barty A, et al. Femtosecond X-ray protein nanocrystallography[J]. Nature, 2011, 470(7332): 73.

## Author contributions

A.X. and X.J. conducted the experiment. A.X. proposed the concept and performed simulation. A.X., W.Y. and Q.Z. analyzed the simulation and experiment data. All authors discussed the results and contributed to writing the manuscript.

## Competing interests

The authors declare no competing interests.

## Data availability

The data that support the plots within this paper and other findings of this study are available from the corresponding author upon reasonable request.

**Correspondence and requests for materials** should be addressed to Q.Z.